\documentclass[12pt]{article}

\usepackage{amsmath}
\usepackage{amssymb}
\usepackage{epsfig}
\usepackage{cite}
\usepackage{color,colordvi}
\usepackage{xcolor}
\usepackage{mathtools}
\usepackage[colorlinks=false]{hyperref}

\newcommand{\kommentar}[1]{}
\DeclarePairedDelimiter\abs{\lvert}{\rvert}%
\DeclarePairedDelimiter\norm{\lVert}{\rVert}%

\makeatletter
\let\oldabs\abs
\def\abs{\@ifstar{\oldabs}{\oldabs*}}
\let\oldnorm\norm
\def\norm{\@ifstar{\oldnorm}{\oldnorm*}}
\makeatother

\begin{document}

\vspace{0.01cm}
\begin{center}
{\Large\bf Memory and the Infrared} 

\end{center}

\vspace{0.1cm}

\begin{center}

{\bf Cesar Gomez}$^{a,b}$\footnote{cesar.gomez@uam.es}, {\bf Raoul Letschka}$^{b}$\footnote{raoul.letschka@csic.es}
\setcounter{footnote}{0}
\vspace{.6truecm}


{\em $^a$Arnold Sommerfeld Center for Theoretical Physics\\
Department f\"ur Physik, Ludwig-Maximilians-Universit\"at M\"unchen\\
Theresienstr.~37, 80333 M\"unchen, Germany}


%


{\em $^b$
Instituto de F\'{\i}sica Te\'orica UAM-CSIC, C-XVI \\
Universidad Aut\'onoma de Madrid,
Cantoblanco, 28049 Madrid, Spain}\\

\end{center}

\begin{abstract}
\noindent
Memory effects in scattering processes are described in terms of the asymptotic  retarded fields. These fields are completely determined by the scattering data and the zero mode part is set by the soft photon theorem. The dressed asymptotic states defining an infrared finite S-matrix for charged particles can be defined as quantum coherent states using the corpuscular resolution of the asymptotic retarded fields. Imposing that the net radiated energy in the scattering is zero leads to the new set of conservation laws for the scattering S-matrix which are equivalent to the decoupling of the soft modes.
The actual observability of the memory requires a non-vanishing radiated energy and could be described using the infrared part of the differential cross section that only depends on the scattering data and the radiated energy. This is the IR safe cross section with any number of emitted photons carrying total energy equal to the energy involved in the actual memory detection. 
 
\end{abstract}
\newpage

Memory effects in gravitational scattering were first discussed in \cite{Zeld} and later on developed in \cite{Christ, Thorne}. Similar memory effects can be derived in classical electromagnetism (see \cite{Wald, Bieri} for a recent discussion). In a series of papers \cite{Strominger, He:2014laa, StromingerLectures} a new understanding of memory effects in connection with soft theorems \cite{Weinberg, Low:1954kd, Low:1958sn} has been put forward. The main result is that the infrared part of the Fourier transform of the classical memory effect is determined by the soft photon or graviton theorems. In addition it was shown in \cite{Stro} that soft photon theorems can be interpreted as Ward identities relative to QED symmetries.

Related with this research the classic topic of infrared divergences in quantum field theories has been revisited, see for instance \cite{Kapec:2017tkm}  (the classic reference where the reader can find most technical details is \cite{YFS}). In theories like QED we have IR divergences due to virtual photons. These divergences can be resummed and regulated. In addition we have soft radiation and we can, at the level of {\it the cross section}, sum over amplitudes for different number of soft emitted photons. These two contributions, namely the one associated with virtual photons and the one coming from summing over different number of final  infrared soft photons lead to infrared divergent pieces that cancel each other in the final {\it cross section}. What remains is an infrared finite cross section. The concrete form of this cross section depends on an {\it infrared scale} $\varepsilon$. More precisely and using the notation of \cite{YFS} the final differential cross section factorizes into a pure infrared part that only depends on the scattering data but not on the details of the scattering and a non-infrared part. This factorization of the cross section depends on the infrared scale used to define the upper limit on the energies for real infrared, measurable photons. The infrared part of the cross section, after infrared divergences are cancelled, depends on the infrared scale $\varepsilon$ in the general form $\frac{\mathrm{d}\sigma}{\mathrm{d}\varepsilon} \sim \text{e}^{G(\varepsilon)}$ with $G(\varepsilon) \sim  \ln\left(\frac{E}{\varepsilon}\right)$. 

The former solution of the infrared problem is not defining an infrared finite S-matrix. The definition of such S-matrix was first addressed in \cite{FK}. 
The key ingredient used in \cite{FK} consisted in modifying the definition of asymptotic states. These are defined using a coherent state dressing operator determined by the asymptotic dynamics.  Using these new asymptotic states for charged particles you can define an IR finite S-matrix for scattering processes with only charged particles in the in and out states. In this context it can be easily observed that for the so-defined IR finite S-matrix  soft photon modes are by construction decoupled. This  FK-decoupling  has been discussed in {\cite{Ware, Porr, Isr, us}. Note that for this scattering S-matrix, with only charged particles in the asymptotic states, conservation of energy implies that no net amount of energy is radiated. This condition leads to a set of conserved charges commuting with S which are equivalent to the decoupling of soft modes. The physical meaning of these symmetries will be discussed in the next section.

In spite of its beauty the S-matrix formalism of \cite{FK} is not directly addressing the most physical discussion on how to deal with real processes where some energy is radiated in the form of real infrared photons. 

How are electromagnetic memory effects related with this issue? The quick answer to this question is the following. In scattering processes among massive charged particles the charges and momentum of the in and out particles determine the non-radiative part of the asymptotic retarded field. These in and out scattering data are enough to extract the zero mode part of the interpolating retarded field and consequently they account for the information contained in the soft photon theorems. In the IR finite S-matrix these zero energy modes are decoupled and moreover they don't lead to any {\it observable} (in a finite amount of time) memory effect.

 The scattering data, although enough to derive the soft photon theorem, are not enough to fix the {\it radiative component} of the retarded field that depends on concrete information on how the scattering process is actually taking place, in particular (in the classical case) on the accelerations. This radiative part of the retarded field carries energy as well as radiative modes with typical frequencies of the order of the inverse of the time scale on which the scattering process is taking place. The {\it observability} of the memory effect using a physical detector depends crucially on this radiated energy. In QED this information is partially encoded in the infrared part of the differential cross section, namely on the dependence on the infrared scale $\varepsilon$ that we can take as equal to the energy involved in the actual detection of the memory effect. In particular we shall associate to memory the infrared part of the cross section that only depends on the scattering data and where we consider an arbitrary number of emitted real infrared photons with total energy  $\varepsilon$ equal to the energy involved in the memory detection.

In this note we shall reduce the discussion of memory to the electromagnetic case and only at the end we will make few comments on similarities and differences with the gravitational case.

\section{Classical memory}
For a given classical scattering where some initial charges $q_j$ with velocities $\mathbf{v_j}$ lead to a final state with charges $q_i$ and velocities $\mathbf{v_i}$ the electromagnetic memory is determined by the retarded field created by the currents $j^{\mu}$ defined by these scattering data. In four dimensions the retarded electromagnetic field at some observation point $O =(\mathbf{ x}, t)$
is given by
\begin{equation}
A^{\mu}({\bf x},t) = \int \frac{j^{\mu}\left({\bf x'} ,t-\frac{r}{c}\right)}{r} \mathrm{d}^3 {\bf x'}
\end{equation}
with $r^2 = (\mathbf {x} - \mathbf {x'})^2$. For small velocities we can Taylor expand the current and define the retarded field as a series in $1/c$.
The field tensor $F_{\mu \nu}$ generated by the moving charges can be expanded in powers of $1/r$. It contains a piece that goes like $1/r^{2}$ that only depends on the velocities of the sources and a piece that goes like $1/r$ that accounts for the radiation emitted during the scattering process.

For an idealised point-like scattering taking place at the origin the radiative part of the retarded field has support on the $u=0$ null hypersurface $t= \frac{r}{c}$. This simply reflects the fact that only at the origin the moving particles entering into the scattering are accelerated. At large distances $\mathbf{x}\gg\mathbf{x'}$ the retarded field is given by
\begin{align}
	A^\mu(x)=\sum_\text{out} \frac{\theta(u)}{r}\frac{q_i v^\mu_i}{1-\mathbf{v_i}\cdot{\hat{\mathbf{x}}}    }+
	\sum_\text{in}  \frac{\theta(-u)}{r} \frac{q_j v^\mu_j}{1-\mathbf{v_j} \cdot \hat{\mathbf {x}}}
\end{align}
where $u$ is determined by the equation $u=t-r$, $\hat{\mathbf {x}}$ is the norm vector of $\mathbf{x}$ and from now on $c=1$.

The field tensor is then given by 
\begin{align}\label{F2}
	F_{\mu\nu}=&\sum_\text{out}\frac{q_i\kappa_{[\mu}v_{\nu]i}}{\kappa_\alpha v^\alpha_i}\left[ \frac{1}{r}
	\frac{\delta(u)}{\kappa_\alpha v^\alpha_i} +
	\frac{1}{r^2}\frac{v_{i\beta} v^\beta_i}{\left(\kappa_\alpha v^\alpha_i \right)^2} \theta(u) \right] \nonumber
	\\ +&\sum_\text{in}\frac{q_j\kappa_{[\mu}v_{\nu]j}}{\kappa_\alpha v^\alpha_j}\left[-\frac{1}{r} \frac{\delta(u)}
	{\kappa_\alpha v^\alpha_j}+\frac{1}{r^2}\frac{v_{j\beta}v^\beta_j}{\left(\kappa_\alpha v^\alpha_j \right)^2}
	\theta(-u) \right]
\end{align}
where
\begin{align}
	\kappa^\mu = \left(1,\hat{\mathbf{x}}\right)  \quad \text{and} \quad v^{\mu}_i=\left(1,\mathbf{v_i}\right)
\end{align}
and indices are raised and lowered in cartesian coordinates by the metric $\eta_{\mu \nu}=\text{diag}(1,-1,-1,-1)$.

As we can see from the former expression the radiative $1/r$ part depends on the concrete classical modelling of the scattering, in this simple case in the form of an instantaneous change of the velocities taking place at the origin. The non-radiative part that goes as $1/r^2$ depends only on the in and out scattering data.

The {\it classical memory} effect associated with a given scattering process where we use as data the in and out four-momenta of the scattered particles is given by the {\it non-radiative fields} $F_\text{in}$ and $F_\text{out}$. However the actual  detection of the memory is determined by the interaction of some charged detector with the interpolating {\it radiative field}. This effect on the memory detector is non-vanishing and {\it observable} due to the fact that the interpolating radiative field carries non-vanishing energy $\varepsilon$. 

\subsection{Spectral resolution}
For future convenience it would be important to work out the spectral decomposition of the asymptotic retarded fields defined by the in and out set of free moving charged particles. The Fourier modes of the retarded field are given by
\begin{align}\label{spectral}
	A^\mu(t,\mathbf{k})= \sum_\text{out} \frac{4 \pi q_i}{\abs{\mathbf{k}}\left(1+\hat{\mathbf{k}} \cdot \mathbf{v_i} \right)} 
	\frac{p_i^\mu \text{e}^{-\text{i}\frac{\mathbf{k \cdot  p_i}}{E_i} t   }}{p_i^\alpha k_\alpha} \Bigg \vert_{t>0}+
	\sum_\text{in}\frac{4 \pi q_j}{\abs{\mathbf{k}}\left(1+\hat{\mathbf{k}} \cdot \mathbf{v_j} \right)} \frac{p_j^\mu \text{e}^{-\text{i}
	\frac{\mathbf{k  \cdot  p_j}}{E_j} t   }}{p_j^\alpha k_\alpha} \Bigg \vert_{t<0}
\end{align} 
where $p_i^\mu$ is the 4-momentum of the $i$th particle and $k^\mu=(\abs{\mathbf{k}},\mathbf{k})$.

The important thing to be noticed is that the Fourier components of the retarded field created by a moving charge with constant velocity $\bf v_i$ are waves with wave vector $\bf k$ but frequency $\omega_i = \bf k\cdot \bf v_i$. These Fourier modes are obviously not real photons with the exception of the soft ${\bf k}=0$ mode. Once we move into quantum field theory these modes will define {\it the quantum constituents} of the coherent state dressing of free moving charged particles.

\subsection{Symmetries, Goldstones and energy conservation}

Classically we can associate with a given scattering process among charged particles the non-radiative retarded fields defined by the in and out scattering data i.e. by the charges, masses and velocities of the incoming and outgoing particles. Let us generically denote $A_\text{in}$ and $A_\text{out}$ these retarded fields. Associated with these data we can formally define a transformation $T: A_\text{in} \rightarrow A_\text{out}$. This transformation is not a gauge transformation since $A_\text{in}$ and $A_\text{out}$, although satisfying the condition
\begin{equation} \label{defLGT}
\lim_{r\rightarrow \infty} r(F_{\mu\nu}(A_\text{in}) - F_{\mu\nu}(A_\text{out})) = 0\text{ ,}
\end{equation} 
have, at order $1/r^2$, different values of the corresponding stress tensor.

Let us now fix the asymptotic kinematical data for the incoming and outgoing charges in such a way that energy and momentum are conserved i.e. $\sum_\text{in}E_{j} =\sum_\text{out} E_{i}$. In this case conservation of energy will imply that the only possible radiated mode is a zero energy zero mode. In classical electrodynamics this constraint is not easy to impose. Indeed if we fix the scattering data and we use those kinematical data to derive the classical radiated field we will only achieve total energy conservation if in addition we take into account the {\it backreaction} of the radiated field i.e. the Abraham-Lorentz forces on the outgoing scattering data. As it is well known this problem cannot be fully solved in classical electrodynamics.

We can however formally impose the conservation of energy on the scattering data for the charged particles which is effectively equivalent to setting the net amount of radiated energy to be equal to zero. To understand the physical meaning of this  {\it zero radiated energy constraint} it can be illustrative
to recall the attempt of Wheeler and Feynman (WF) \cite{Wheeler:1945ps, Wheeler:1949hn} to define in classical electrodynamics the radiative reaction on sources in the context of the absorber theory. Indeed if we think that all the radiation emitted is absorbed leading to zero radiated energy we get that asymptotically we can impose the WF condition:
\begin{equation}\label{matching}
F_\text{ret}^{\mu\nu} = F_\text{adv}^{\mu\nu}
\end{equation}
for the radiative part of the total advanced and retarded fields.\footnote{Note that the condition (\ref{matching}) allows us to define the WF field associated with a moving charge as $1/2(F_{ret} + F_{adv})$.} Generically, although in Maxwell theory we have the advanced and retarded solution, only the retarded part of the radiative field is actually considered as physical. Thus the former condition makes sense if we have a formal {\it absorber} and no net radiation carrying non-zero energy  is left unabsorbed. 

In scattering language we can think of the advanced field as associated with some incoming radiation and the retarded field as the outgoing radiation, so if we consider a scattering with in state defined by a set of {\it only} charged particles (and zero radiation) the former condition \eqref{matching} only makes sense for the zero mode part that does not carry any energy. 

The equality between retarded and advanced fields \eqref{matching} leads to a set of conservation laws where the classical charges can be defined by the convolution of \eqref{matching} with arbitrary test functions \cite{Campiglia:2017dpg}. The so called soft charges can be defined as those determined by the zero mode part of the retarded and advanced fields. 

In a scattering process among charged particles where we use as scattering data a set of in and out momenta for the charged matter satisfying conservation of energy i.e. with no net radiation, we can impose the condition \eqref{matching} and these charges will act as symmetries of the S-matrix. Since there is no radiation the only relevant piece is the zero mode soft part. In this case any memory effect defined as the difference between the non-radiative part of the retarded fields created by the incoming and outgoing particles is physically {\it unobservable}. This unobservability becomes equivalent, in the S-matrix language, to the decoupling of the radiative zero mode.

In summary the ``new symmetries" of the QED S-matrix \cite{Stro, Kapec:2015ena} are a consequence of imposing what we can call the WF condition or in more physical terms, the absence of any loss of energy in the form of radiated infrared photons. This condition is naturally implemented in any S-matrix formulation where in and out states are sets of charged particles. However in order to have observable memory effects a certain amount of energy should be radiated and in that case we need to work with the differential cross section. Once some energy is actually radiated  we cannot impose \eqref{matching} since this energy is only contained in the retarded part of the field. 

In reality the probability that in a physical scattering we have zero net radiated energy is indeed zero, so these symmetries of the IR finite S-matrix only account for the soft theorem part. We can think of the symmetries for zero energy radiation processes as being spontaneously broken with the $\mathbf{k}=0$ soft mode as a Goldstone boson. However we would like to stress that whenever we have a real amount of energy radiated with no incoming radiation, which is actually always the case, the condition \eqref{matching} can only be imposed for the zero mode part which is what, as we shall discuss in moment, you actually do in the definition of the IR finite S-matrix.

\section{Memory and infrared QED}
\subsection{IR finite S-matrix}
As was already pointed out in \cite{FK} a prescription to define an IR finite S-matrix was partially developed. The key ingredient in this construction was to use the asymptotic dynamics in order to define new asymptotic states by dressing standard Fock matter states $|i\rangle$ with the coherent state of photons sourced by the {\it asymptotic current} $J_{as}^{\mu}$. We can represent this dressing as
\begin{equation}
|i\rangle \rightarrow  \text{e}^{R (J_{as}^{\mu})} |i\rangle
\end{equation}
We can now easily identify the operator $R$. Using the spectral decomposition of the retarded field (\ref{spectral}) created by the asymptotic free moving charges we can define the quantum resolution\footnote{For other examples of the same technique see \cite{Dvali:2015jxa} and \cite{Dvali:2017eba}.} of this field using as quantum constituents, quanta with momentum $\mathbf{ k}$ and frequency
\begin{equation}
\omega_i = \mathbf{ k \cdot v_i} \text{.}
\end{equation}
Denoting the creation and annihilation operators for these quanta $b_\mathbf{k}$ and $b^{\dagger}_\mathbf{k}$ the corresponding coherent state will be defined by the operator
\begin{equation}
\text{e}^{\int \mathrm{d}^3k \left(\sum \frac{4 \pi q_i}{\abs{\mathbf{k}}\left(1+\hat{\mathbf{k}} \cdot \mathbf{v_i} \right)} 
	\frac{p_i^\mu \text{e}^{-\text{i}\frac{\mathbf{k \cdot p_i}}{E_i} t   }}{p_i^\alpha k_\alpha} b^{\dagger}_{\mathbf{ k}\mu} \right) }
	\end{equation}
	acting on the vacuum defined by $b_\mathbf{k}|0\rangle =0$. If we want to use the creation and annihilation operators $a_\mathbf{k}$ and $a^{\dagger}_\mathbf{k}$ of the Fock space of free photons with dispersion relation $\omega= \abs{\bf k }$ we need to transform $b$ modes into $a$ photons. This leads to the FK expression derived from the asymptotic dynamics, namely
\begin{equation}
\text{e}^{ \int \frac{\mathrm{d}^3 k}{k_0} \frac{p^\mu}{p_\alpha k^\alpha} \text{e}^{i \left(\frac{p_\alpha k^\alpha}{p_0}\right)t} a_{\mathbf{k}\mu}^{\dagger} }\text{.}
\end{equation}
By construction on these coherent states the expectation value of the field operator $\hat A$ is given by the classical retarded field. Note that these coherent states contain an infinite number of ${\bf k}=0$ photons. If in the scattering process we impose zero energy radiated then the total number of modes in the in and out states will be conserved.

We can consider a more complicated coherent state of photons describing the whole radiative part of the retarded field and to think of this coherent state as a sort of {\it domain wall} interpolating between the asymptotic in and out retarded fields. The soft photon theorem accounts for the zero mode part of this {\it domain wall}. The radiated energy acting on the potential memory detector is roughly what we can interpret as the {\it mass} of this photonic domain wall. 

The IR finite S-matrix is defined by 
\begin{equation}
\lim _{t\rightarrow \infty}\text{e}^{R(-t)^{\dagger}} S \text{e}^{R(t)}
\end{equation}
where $\text{e}^{R(t)}= \text{e}^{ \int dp \frac{\mathrm{d}^3 k}{k_0} \frac{p^\mu}{p_\alpha k^\alpha} \left( \text{e}^{i\left(\frac{p_\alpha k^\alpha}{p_0}\right)t} a_{\mathbf{k}\mu}^{\dagger} -h.c \right)\rho(p) }$ for $\rho(p) = \sum_i \delta(p-p_i)$. This S-matrix
satisfies the decoupling of soft modes \cite{Porr, Isr, us}\footnote{For a more rigorous proof see \cite{Mischa}.}
\begin{equation}
\lim_{\mathbf{k}\rightarrow0} [S,a_\mathbf{k}] =0\text{ .}
\end{equation}
The so-defined S-matrix is IR finite due to the fact that the former dressing factor cancels the infrared divergences (after resummation) coming from the virtual photon self energies.

Note that in this S-matrix we are imposing the zero energy radiation condition \eqref{matching} and consequently the S-matrix commutes with the charges defined by convoluting \eqref{matching} with arbitrary test functions. These Ward identities are simply reflecting the kinematical constraints we are imposing on the scattering states, namely vanishing net energy in the form of radiation for in states without real photons and are fully equivalent to the decoupling of soft photons. Note also that, in this case, the so-called hard charges \cite{Stro, Kapec:2015ena, us} are absorbed in the dressing. 

It is important to stress that the decoupling of soft modes should not be confused with the absence of {\it observable} memory effects. Indeed as already stressed observable memory requires a certain amount of energy in the retarded field to be radiated in the form of infrared emitted photons and therefore does not satisfy the S-matrix matching condition for the charged kinematical data.

\subsection{QED measure of memory}

Given a scattering process in QED we can associate, as a way to characterize the memory, the differential cross section $\frac{\mathrm{d}\sigma}{\mathrm{d}\varepsilon}$ for $\varepsilon$ the radiated energy in the form of infrared photons.  The dependence of the cross section on $\varepsilon$ is well known in QED \cite{YFS}. We shall be interested only in the infrared part of the cross section i.e. in the part that only depends on scattering data. 
\begin{equation}
\frac{\mathrm{d}\sigma}{\mathrm{d}\varepsilon}\bigg \vert_\text{IR} \sim A \text{e}^{\ln \frac{E}{\varepsilon}}
\end{equation}
with $A$ being a finite coefficient depending only on the scattering data. This infrared part of the cross section corresponds to having arbitrary number of emitted infrared photons with total energy equal or less than $\varepsilon$, i.e. it is the cross section $\sigma(2 \rightarrow 2 + soft (E_\text{soft} \leq \varepsilon))$ and is an IR safe quantity.

     The important message of these cross sections is the dependence on the energy radiated. This is important for understanding the real nature of the memory. In fact we could think of nullifying the memory by pushing $\varepsilon\rightarrow0$. In this case the only remnant will be the zero mode part of the radiative mode that  is actually decoupled. However the dependence of the cross section on $\varepsilon$ is telling us that such a formal limit cannot be taken or equivalently that the actual probability to scatter without radiating is zero. The interpolating radiative field measured by the memory detector contains energetic modes in addition to the Goldstone zero mode piece. The actual interaction of the detector with these modes is what makes the memory effect, in scattering processes, actually observable in a finite amount of time.

\section{Summary and some comments on the gravitational memory}

To summarise we have observed that although in an IR finite S-matrix of the type discussed in \cite{FK} the Goldstone part of the radiative mode is decoupled this does not nullify electromagnetic memory effects. These effects are classically due to the non-vanishing energy carried by the radiative part of the retarded field created by the scattering process and show up quantum mechanically already in the infrared dependence  of the differential cross section on the amount of energy radiated in the form of real infrared quanta.
In this sense memory is intimately connected with the {\it infrared scale} appearing in the standard computation of differential cross sections in QED. Pushing formally this IR scale to zero and working with the IR finite S-matrix satisfying the condition \eqref{matching} nullifies memory as an observable effect. Note also that memory accounts for the radiative {\it backreaction} on the outgoing charged particles of the radiated energy.

Regarding gravity the discussion of classical gravitational memory is formally identical to the electromagnetic case.  Again in this case {\it observable} memory effects are associated with the non-zero mode part of the interpolating radiative field. In quantum field theory language we can extract the dependence of the differential cross section on the infrared scale as we do in the electromagnetic case. This problem has been recently discussed for graviton scattering in \cite{Ven}. This construction can be used to define IR safe quantities to be associated with gravitational memory.

A very attractive idea discussed in \cite{Str, HPS, HPSsuperrotation} suggests a connection between the gravitational memory and the information paradox. The basic point of the idea is to extend the gravitational memory associated with radiation going through the null infinity to the analog problem for the black hole horizon itself. This extension is by no means straightforward for a simple reason. While in the case of the null infinity the radiated energy (involved in the memory) interpolates between two Minkowski asymptotic metrics (related by a supertranslation) in the case of the horizon any gravitational memory interpolates between two black hole metrics with {\it different mass}. The zero mode part of this interpolating metric  can be formally used to define {\it soft hair}. However this zero mode part is effectively decoupled and unobservable \cite{us, Porr}. The conservation laws associated with the gravitational generalization of \eqref{matching} can only deal, even in the presence of horizons, with the zero energy part of the involved radiative modes and consequently does not lead to any observable effect.\footnote{From a purely historical point of view an interesting discussion on WF theory and gravity can be found in \cite{Wheeler:1945ps}} The observable gravitational memory, as it is the case with the electromagnetic memory, is a radiative backreaction effect that can be only worked out  quantum mechanically, in other words the zero mode part which corresponds to no radiative backreaction (and can be described in purely classical or semiclassical terms) is ineffective to solve the information paradox. The quantum backreaction, as well as the relevant quantum hair, is $O(1/N)$ for $N$ the black hole entropy \cite{Dvali:2012rt, Averin:2016hhm}.

However, for a scattering process where some initial state leads to the formation of a black hole that, by complete evaporation, leads to a pure final state of radiated particles, we can ask, for the corresponding cross section, about the infrared gravitational memory. This is a question about how much initial energy is radiated in the form of infrared gravitons (i.e. with energies smaller than the IR scale). Part of the information on how the whole process unitarizes could be encoded in these modes. Information about the radiated energy is contained in the asymptotic non-radiative retarded fields.   A natural question is if in the case of gravity there exist a natural IR scale. A possible answer is to use the gravitational radius of the center of mass energy. A related discussion can be found in \cite{Ven} in connection with the classicalization $2\rightarrow N$ approach \cite{Dvali:2014ila} to black hole formation. We hope to come back to this issue in a future publication.

\section*{Acknowledgements}
We would like to thank Gia Dvali for insightful comments related to the observability of memory and Kepa Sousa and Mischa Panchenko for many discussions and comments.

The work of C.G. was supported in part by Humboldt Foundation and by Grants: FPA 2009-07908 and  ERC Advanced Grant 339169 "Selfcompletion''. The work of R.L. was supported by the ERC Advanced Grant 339169 "Selfcompletion''.


\end{document}